\documentstyle[prd,aps,eqsecnum,tighten]{revtex}
\begin{document}
\draft
\title{On the Concept of Spin} 
\author{ I.~B.~Pestov\thanks{Electronic address:
pestov@thsun1.jinr.ru}}
\address{Bogoliubov Laboratory of Theoretical Physics, Joint
Institute for
Nuclear Research \\ Dubna, 141980, Russia}

\date{\today}
\maketitle
\begin{abstract}
It is substantiated that spin is a notion associated with the group
of internal symmetry that is tightly connected with the geometrical
structure of spacetime. The wave equation for the description of a
particle with spin one half is proposed. On this ground it is shown
that the spin of electron is exhibited through the quantum number and
accordingly the Dirac equation describes properties of particles
with the projection of spin  $\pm \hbar /2.$ On the contrary, we put
forward the conjecture that the spin of the quark cannot be
considered as a quantum number, but only as an origin of a
non-abelian gauge field.  The reason is that the quark and electron
>from physical, geometrical and group-theoretical points of view
differ from each other. It is a deep reason for understanding
quark-lepton symmetry and such important phenomena as quark
confinement.

\end{abstract}
\pacs{12.20.-m, 12.20.Ds, 78.60.Mq}

\section{Introduction} 
This report is devoted to the new approach to the problem of 
theoretical description of the particle with spin one half and 
especially in the context of the theory of leptons and quarks. As it 
is well known, spin is a fundamental notion of modern physics and its 
meaning has tendency to grow with time. But what stands out is the 
absence of spin one half operators $S_1, \, S_2, \, S_3$  with the 
properties \begin{equation} [S_i, S_j ] = i e_{ijk} S_k 
\end{equation} \begin{equation} S_1^2 + S_2^2 + S_3^2 = S^2 = 
\frac{3}{4}= s(s+1) \end{equation} \begin{equation} [H,\, S_i ] = 0, 
\end{equation} in the Dirac theory of electron, where $H$ is the 
Hamiltonian proposed by the Dirac. As a consequence of nonvalidity of 
the last relation in the framework of the Dirac theory the 
corresponding tensor of spin angular momentum is not conserved.  
Description of the spin in the framework of the Pouncare group is not 
satisfactory because by this method we can derive important 
information on the state of a field that corresponds to a particle in 
question but not on its internal property such as spin is. In view of 
this, for the description of a particle with spin one half, a new 
wave equation is proposed below which verifies that equations 
(1),(2),(3) are fulfilled.

 \section{On the wave equation for a particle \protect{\\} with spin
one half}

Our proposal is the following system of equations for the
description of particles with spin $1/2$
\begin{equation}
    P^i{\psi_i} = \frac{mc}{\hbar} {\psi},
\end{equation}
\begin{equation}
P_i {\psi}_j - P_j
    {\psi}_i - i e_{ijkl} P^k {\psi}^l =\frac{mc}{\hbar} {\psi}_{ji},
\end{equation}
\begin{equation}
P^i{\psi}_{ij} - P_j {\psi} = \frac{mc}{\hbar} {\psi}_j.
    \end{equation} with $$P_i = \nabla_i - (ie/\hbar c) A_i,$$ where
	 $A_i$ is the vector potential of an electromagnetic field,
 $\nabla_i$  is the covariant derivative with respect to the Levi-
 Civita connection $$ \Gamma^i_{jk} =  \frac{1}{2}g^{il}(\partial_j
 g_{kl}+\partial_k g_{jl}-\partial_l g_{jk}  ),$$ $\psi$ is a scalar
 field, $\psi_{i}$ is a vector field and $\psi_{ij}$ is a self-dual
 bivector $$ i \psi_{ij} = \frac{1}{2}
e_{ijkl}{\psi}^{kl}= \tilde{\psi}_{ij}.
$$
Completely antisymmetric Levi-Civita tensor $e_{ijkl}$ is normalized
as follows $e_{0123} = \sqrt{-g},$ where $g$ is the determinant of the
metric tensor. Thus, the wave function $$\Psi = (\psi, \, \psi_i, \,
\psi_{ij})$$ has eight components. We consider the general case of
curved spacetime with metric $ds^2 = g_{ij}dx^i dx^j$ since in our
approach its geometrical properties are tightly connected with the
concept of spin as a fundamental quantum number.

Spin one half group is the symmetry group of the wave equation and it
is defined as follows
$$ \psi \Rightarrow \psi'= - \frac{1}{4} {\Sigma}_{ij} {\psi}^{ij}$$
$$ \psi_{ij} \Rightarrow \psi'_{ij} = \frac{1}{2} \Sigma_{ik}\psi^k_j
 - \frac{1}{2} \Sigma_{jk}\psi^k_i+ \Sigma_{ij} \psi,$$ $$\psi_i
\Rightarrow \psi' = \Sigma_{ik}\psi^k, $$ where $\Sigma_{ij}$  is a
self-dual bivector, $i \Sigma_{ij} = (1/2) e_{ijkl}\Sigma^{kl} =
\tilde \Sigma_{ij}.$ It is easy to recognize the origin of this group
of transformations when the equations of
second order are derived from the (4),(5),(6).  These eqs. are of the
form $$(P^iP_i + \frac{m^2c^2}{\hbar^2}) \psi_j = \frac{ie}{\hbar c}
H_{jk}  \psi^k,$$ $$(P^iP_i +
\frac{m^2c^2}{\hbar^2}) \psi = \frac{ie}{4\hbar c} H_{jk} \psi^{jk}, $$ $$(P^iP_i + \frac{m^2c^2}{\hbar^2}) \psi_{jk} =
\frac{ie}{2\hbar c} (H_{jl}\psi^l_k - H_{kl}\psi^l_j)-\frac{ie}{\hbar
c} H_{jk} \psi, $$ where $ H_{jk}= F_{jk} - i \tilde
F_{jk}, \quad    F_{ij} = \partial_i A_j - \partial_j A_i$ is the
bivector of the electromagnetic field.

However, the transformations of the spin
one half group act in the space of solutions of the wave equation only
under the condition  \begin{equation} \nabla_i
\Sigma_{jk} = 0. \end{equation}   This equation is very
important because it connects the spin with the geometrical structure
of spacetime. In the flat spacetime with Minkowski metric equation
(7) has three linear independent solutions which define the operators
$S_1, \, S_2, \, S_3$  with the necessary properties
 $$[S_i, \, S_j] = i e_{ijk} S_k, \quad S_1^2 + S_2^2 + S_3^2 =
 \frac{3}{4}, \quad [H,S_i] = 0,$$ where $H$ is the Hamiltonian of
 the wave equation in question.  Thus, in the Minkowski spacetime
 the spin one half symmetry can be considered as global  and we
 have no nonformal reason to introduce the gauge field corresponding
 to this symmetry in spite of the fact that it is internal.

 Let  $\Psi_{\pm}$  be the wave functions that satisfy the
 equations
 $$S_3 \Psi _{\pm} = \frac{1}{2} \Psi _{\pm}.$$
It can be shown that $\Psi_{+}$ ( or $\Psi_{-}$ ) is equivalent to
the Dirac wave function in the sense that wave equation for
$\Psi_{+}$ is equivalent to the Dirac equation. So, the Dirac equation
describes properties of particles with the projection of spin
$\pm{\frac{1}{2}}.$  Now it is clear why the tensor of spin angular
momentum is not conserved in the Dirac theory.

  \section{On the wave equation for quarks }

Equation (7)  has no solutions in the spacetime of
constant curvature. This follows from the integrability conditions of
this equation and the main property of  spacetime manifolds
of that kind.  Thus, in the spacetime of constant curvature the spin
one half group can be realized only as a group of local symmetry.
It means that in the case in question spin is the origin of the
Yang-Mills fields and the Planck constant characterizes the
strength of interactions. We should not introduce a special
constant of interaction because in the case of non-abelian gauge
group there is no gauge-invariant conserving quantity like
the electric charge in electrodynamics.

Now it is natural to put forward the idea that in the realm of the
strong interactions spacetime geometrically can be represented as
a one-sheeted hyperboloid
    $$ (x^0)^2- (x^1)^2- (x^2)^2- (x^3)^2 - (x^4)^2 =  - a^2,$$
in the 5d Minkowski spacetime with coordinates
   $x^0, x^1, x^2, x^3,x^4.$  Here $a$  is a constant which can be
interpreted geometrically as the radius of 3-dimensional sphere,
considered here as a space section, and  physically as the size of
region of quark confinement.  We here treat the quark as a pointlike
particle in the space of constant positive curvature $S^3.$  For
comparison of leptons and quarks we note that free motion of
the electron is represented as a straight line in the Euclidean usual
space and free motion of the quark is a circumference  on the 3d
sphere.  This correlation between leptons and quarks will be
continued with an example of the Coulomb law for these
objects.  As it is well known, the electron Coulomb potential
$$\phi_{e}(r) = \frac{e}{r}$$ is the fundamental solution of the
Laplace equation $\triangle {\phi} = \mbox{div\, grad} \,{\phi} = o.
$ In accordance with our conjecture the Coulomb potential for quarks
can be derived as follows. Consider the stereographic projection
$S^3$ from point (0,0,0,-a) on the sphere $x^2+y^2+z^2 \leq a^2:$
$$x^1 = fx, \quad x^2 = f y, \quad x^3 = f z, \quad x^4 =
a(1-f),$$ where $f= 2a/(a^2 + r^2).$  Then, it follows that the
element of length on the 3d sphere can be represented in the form $$
ds^2 = f^2 (dx^2 + dy^2 + dz^2)$$ and hence the Laplace equation on
$S^3$ can be written as follows $$\triangle \phi= f^{-3} \mbox {div}
 (f \mbox{grad }\, \phi) = 0.$$ We look for the solution to this
 equation that is invariant under the transformation
 group $SO(3)$ with generators $$x^1 \frac{\partial}{\partial x^2} -
 x^2 \frac{\partial} {\partial x^1} =  x \frac{\partial}{\partial y}
 - y \frac{\partial}{\partial x} , etc.   $$   This subgroup of the
   $SO(4)$  group is determined by fixing the point $(0,0,0,-a).$ Let
   $$\psi = f \frac{1}{r} \frac{d\phi}{dr}.$$  Since $$\triangle \phi
= f^{-3}(r \frac{d\psi}{dr} + 3\psi) =  r^{-2} f^{-3}\frac{d}{dr}(r^3
   \psi),$$  then $r^3\psi = c_1 = constant.$ Thus, $$
\frac{d\phi}{dr} = c_1 \frac{a^2 + r^2}{2a^2r^2} = c_1
   (\frac{1}{2r^2} + \frac{1}{2a^2})$$  and hence
   \begin{equation}
		 \phi_q = c_1 ( -\frac{1}{2r} + \frac{r}{2a^2}) + c_2.
   \end{equation}
Generally speaking, expression (8)  we have derived coincides with
the well-known Cornell potential [1],[2]. If we demand that $\phi_e(a)
= \phi_q(a),$ then $c_2= e/a $ and the Coulomb law for quarks has
the form \begin{equation} \phi_q(r) = q(\frac{1}{2r} - \frac{r}{2a})
			   + \frac{e}{a}, \end{equation} where $q$ is the
quark charge. From this consideration it follows that equations
(4)-(6) can be considered as basic equations for quarks in
the new geometrical framework described shortly above.

\section{Conclusion}

Let us summarize the results obtained and the problem to be solved.
Now we can really avoid the introduction of mysterious and
artificial concept of "isotopic space" in the realm of high energy
physics if the spin is really a fundamental concept. It is
shown that the interquark potential expresses the Coulomb law for
quarks and, in fact, coincides with the well-known Cornell
potential that was first very successfully used by the Cornell group.
At large $a,$ when $a \rightarrow \infty,$ from the theory of
quarks one can deduce the theory of electrons but with
electrons evidently deconfined, because in this case the
region of confinement is the 3d Euclidean space. Thus, the symmetry
between quarks and leptons has a natural explanation.  Equation (9)
can also be considered as a modification of the Coulomb law on
short distances. Wave equations (4)-(6) describe not only the
spin but the fine structure of the hydrogen atom.  This is not
surprising because of the connection with the Dirac equation. It is
of some interest represent that solutions may be written explicitly
by using only spherical vector harmonics.  In the context of quark
hypothesis it is very important  to investigate equations (4)-(6)
for the case of 3d sphere in more detail. Much has to be done, but
the problem is worth efforts as many interesting applications
become possible.

\end{document}